\documentclass[sigconf]{acmart}
\newcommand{\todo}[1]{#1}

\AtBeginDocument{%
  \providecommand\BibTeX{{%
    \normalfont B\kern-0.5em{\scshape i\kern-0.25em b}\kern-0.8em\TeX}}}

\copyrightyear{2021} 
\acmYear{2021} 
\setcopyright{acmcopyright}
\acmConference[ASP-DAC 2021]{ACM Asia South Pacific Design Automation Conference}{Tokyo}{Japan}

\acmBooktitle{ASP-DAC 2021: ACM Asia South Pacific Design Automation Conference, June 18--21, 2021, Tokyo, Japan}
\acmPrice{15.00}
\acmDOI{10.1145/3394885.3431635}
\acmISBN{978-1-4503-7999-1/21/01}

\begin{document}

\title{Uncertainty Modeling of Emerging Device based Computing-in-Memory Neural Accelerators with Application to Neural Architecture Search}

\author{Zheyu Yan}
\email{zyan2@nd.edu}
\affiliation{%
  \institution{University of Notre Dame}
}
\author{Da-Cheng Juan}
\email{x@dacheng.info}
\affiliation{%
  \institution{National Tsing Hua University }
}

\author{Xiaobo Sharon Hu}
\email{shu@nd.edu}
\affiliation{%
  \institution{University of Notre Dame}
}

\author{Yiyu Shi}
\email{yshi4@nd.edu}
\affiliation{%
  \institution{University of Notre Dame}
}

\renewcommand{\shortauthors}{Yan, Juan, Hu and Shi}

\begin{abstract}
Emerging device-based Computing-in-memory (CiM) has been proved to be a promising candidate for high-energy efficiency deep neural network (DNN) computations. However, most emerging devices suffer uncertainty issues, resulting in a difference between actual data stored and the weight value it is designed to be. This leads to an accuracy drop from trained models to actually deployed platforms. In this work, we offer a thorough analysis of the effect of such uncertainties-induced changes in DNN models. To reduce the impact of device uncertainties, we propose UAE, an uncertainty-aware Neural Architecture Search scheme to identify a DNN model that is both accurate and robust against device uncertainties.
\end{abstract}






\maketitle

\section{Introduction}
Deep Neural Networks (DNNs) have achieved superhuman performance in various perception tasks and have become one of the most popular solutions for these applications. Thus, there is an obvious trend in deploying DNNs on edge devices such as automobiles, smartphones, and smart sensors. However, implementing computational intense DNNs directly on edge devices is a significant challenge due to the limited computation resource and constrained power budget of these devices. Moreover, most of the DNN accelerator designs are confined in a design space where the researchers only consider conventional von-Neumann architectures (e.g., GPUs, mobile CPUs, or FPGAs) as candidate platforms. In von-Neumann architectures, data movement inevitably becomes the bottleneck for system efficiency, due to the well-known memory wall where the computational unit must fetch and store data from the memory hierarchy.

\begin{figure}[t]
\begin{center}
\centerline{\includegraphics[trim=0 140 300 0, clip, width=0.8\linewidth] {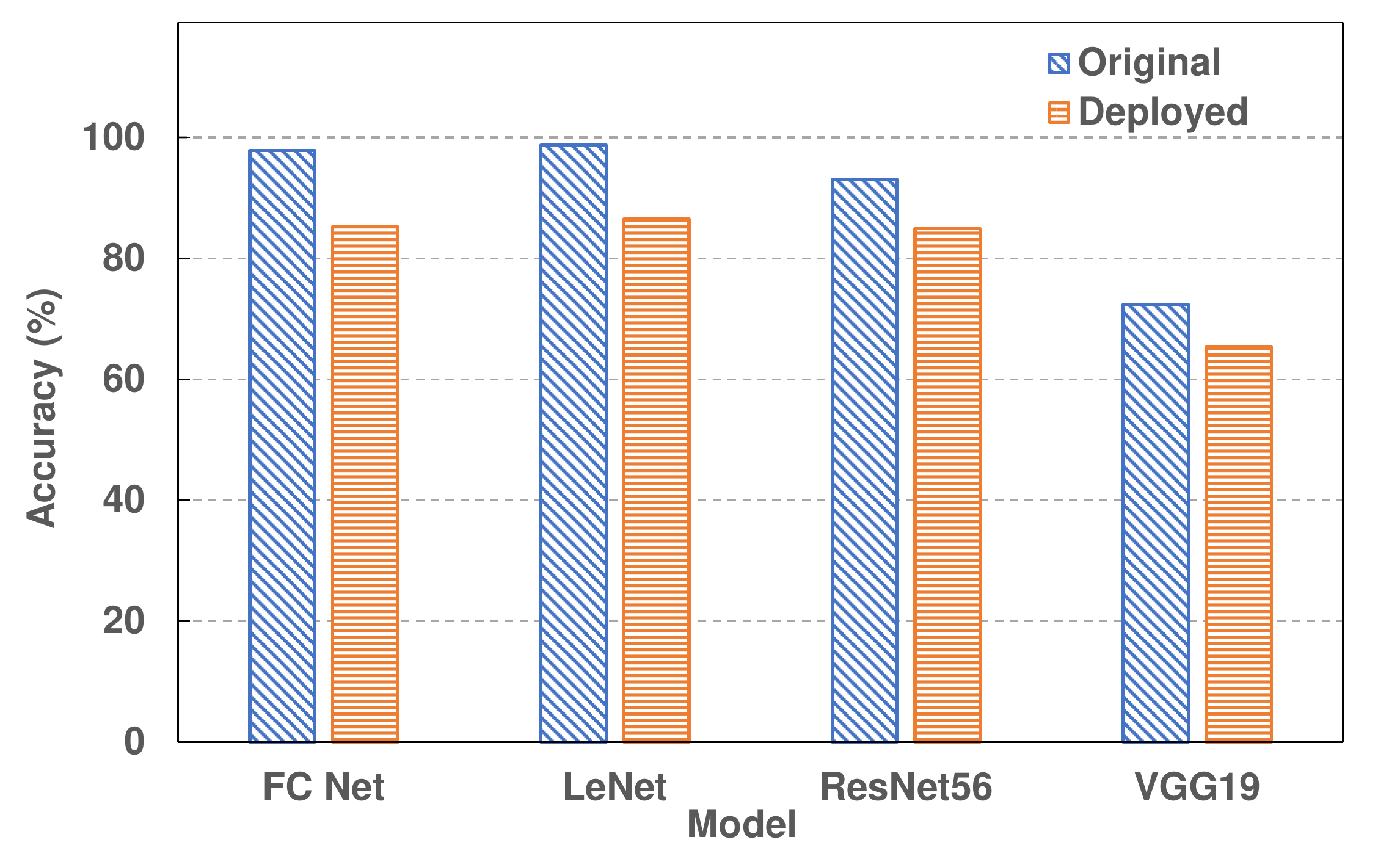}}
\caption{Accuracy differences between models trained in data center and deployed on CiM simulations for different neural architectures. FC Net and LeNet target MNIST, ResNet 56 targets CIFAR-10 and VGG-19 targets ImageNet. An accuracy drop close to 10\% can be observed.}
\label{fig:int}
\vspace{-0.5cm}
\end{center}
\end{figure}

Emerging device-based Compute-in-Memory (CiM) neural accelerators~\cite{ielmini2018memory} offer a great opportunity to break the memory wall with special architectural advantages. CiM architectures offer reduced data movement by in-situ weight data access~\cite{sze2017efficient}. \todo{Highly efficient emerging devices (\emph{e.g.} RRAMs, STT-RAMS, and FeFETs) can be devised to offer higher energy efficiency and higher memory density compared with traditional MOSFET~\cite{shafiee2016isaac} based designs.} However, such accelerators suffer greatly from design limitations. Non-ideal conditions of emerging devices due to their non-ideal manufacturing process induce uncertainties on emerging devices. These uncertainties, such as device-to-device (D2D) variations, thermal noise, and retrieval limitations, cause value changes that, the weights in the actually deployed accelerators may be different from the desired weight value trained offline in data centers. This weight value change leads to performance degradation in actual accelerator implementations. As an illustration, we train the four models, multilayer perceptron (MLP) and LeNet for MNIST, ResNet 56 for CIFAR-10, and VGG-19 for ImageNet, to state-of-the-art accuracy and deploy them on CiM simulation tools~\cite{feinberg2018making}. As shown in Fig.~\ref{fig:int}, an accuracy degradation of close to 10\% is observed in each model implementation.

The device uncertainty-induced performance degradation has been studied from different perspectives, including device-level observations~\cite{zhao2017investigation}, architecture level analysis~\cite{jiang2020device}, and behavioral level explorations~\cite{yan2020single}. Finding suitable pairs of DNN models and hardware designs that can together offer both desirable hardware reliability and high inference accuracy requires great effort.

Neural Architecture Search (NAS)~\cite{zoph2016neural, zoph2017learning, zeng2020towards} is one of the most successful efforts to address this issue. NAS liberates human labor from endlessly exploring optimal handcrafted DNN models by automatically identifying neural architectures that can offer desired performances from a pre-determined search space. Co-exploration of neural architecture and hardware design~\cite{jiang2019accuracy, jiang2020standing, jiang2020hardware} pushes this concept further by incorporating hardware design specifications into NAS search spaces, so as to offer neural architecture-hardware design pairs that are accurate, efficient, and robust against hardware uncertainties.

In this work, we adopt a statistical analysis perspective to study the effect of device uncertainties on the performance of DNNs. We model the emerging device uncertainty as a whole into Gaussian noise on weights and thoroughly investigate the behavior of different DNN models under such uncertainties. We conduct a Monte-Carlo simulation-based analysis on the statistical behavior of the models under the influence of device uncertainties. We then abstract our analysis results to offer supports for NAS applications. The detailed contributions of this work are:

\begin{itemize}
    \item We propose a Monte-Carlo simulation-based experimental flow to measure the device uncertainty-induced perturbations to DNN models.
    \item We then thoroughly investigate the behaviors of different DNN models under such perturbations and show that the value changes of their output vectors follow Gaussian distribution.
    \item To alleviate this effect, we then propose UAE, a device uncertainty-aware NAS framework, to search for architectures that are more robust to device uncertainties.
\end{itemize}

Experimental results show that UAE offers a 2.49\% higher accuracy than NACIM~\cite{jiang2020device} with 1.2x of time consumption. By further increasing search complexity, UAE reaches 6.39\% higher accuracy than NACIM with 2.5x of search time.
\section{Background}
\subsection{CiM DNN Accelerators}
Researchers have proposed different crossbar-based CiM DNN accelerator architectures~\cite{shafiee2016isaac, chi2016prime} for efficient DNN inference. We assume an ISAAC-like architecture~\cite{shafiee2016isaac}, and the architecture of the system is organized into multiple tiles, with crossbar arrays as the heart of each tile. 
The crossbar not only stores the synaptic weights but also performs dot-product computations. One certain crossbar is dedicated to processing a set of neurons in a given DNN layer. The outputs of that layer are fed to other crossbars that are dedicated to
processing the next layer. The computation in the crossbar is performed inanalog
domain. However, ADC and DAC are used to convert the signal from and to the analog domain dot-product computation and other digital domain operations needed in DNN computation. 

\begin{figure}[h]
\begin{center}
\centerline{\includegraphics[trim=0 150 550 0, clip, width=0.6\linewidth] {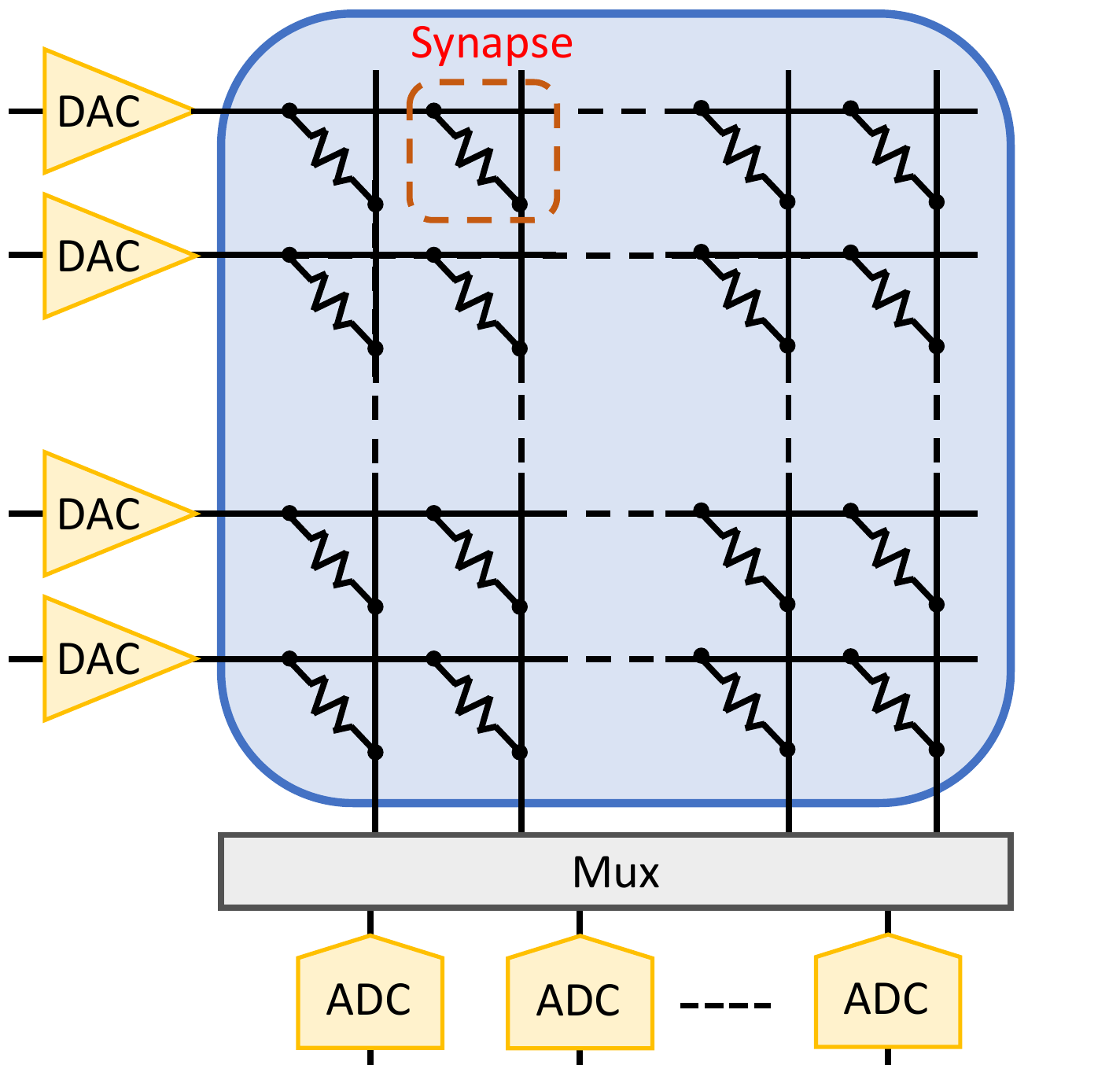}}
\caption{Illustration on crossbar architecture.}
\vspace{-0.5cm}
\label{fig:crossbar}
\end{center}
\end{figure}

Crossbar is the key component of CiM DNN accelerators. As shown in Fig.~\ref{fig:crossbar}, a crossbar can be considered as a processing element for matrix-vector multiplication where matrix value (\emph{i.e.} weights for NNs) are stored at the cross point of each vertical and horizontal line with resistive emerging devices such as RRAMs and FeFETs, and each vector value is propagated through horizontal data lines. In this work, we assume an RRAM-based design. The calculation in crossbar is performed in analog domain but additional peripheral digital circuits are needed for other key NN operations (\emph{e.g.}, non-linear activation), so DAC and ADCs are adopted between different components.

Device-level limitations confine the application of crossbars. The precision of ADC and DACs limits the precision of DNN activations for each layer and the non-ideal characteristics of emerging devices impose noises on the weights of deployed DNNs.

\subsection{Device Variation}

In this work, we assume an RRAM-based crossbar design. RRAM devices suffer various types of faults due to manufacturing and runtime non-idealities. Noise sources that are directly relevant to crossbar-based designs include thermal noise, shot noise, random telegraph noise (RTN), and programming errors~\cite{feinberg2018making}. When the circuitry is used for inference, programming errors due to device-to-device variations could be the dominant error source.

Write-and-verify~\cite{alibart2012high, niu2012low, xu2013understanding} is a simple, accurate, and widely used programming scheme for RRAMs. The key operation is to iteratively apply a series of short pulses and check the difference between current and target resistance, converging progressively on the target resistance. In deploying accelerators for Neural Network inference, this time-consuming progress is tolerable because once programmed, no more modifications to the resistance are needed during the entire life span of the accelerator. Although this scheme pulls down the D2D variation-induced error to less than 1\%, a significant error drop can still be observed in conditions shown in Fig.~\ref{fig:int}.




\subsection{Neural Architecture Search}
Neural Architecture Search (NAS) has achieved state-of-the-art performance in various perceptual tasks, such as image classifications~\cite{yang2020co_1, yang2020co_2}, inference security~\cite{bian2020nass} and image segmentation~\cite{yan2020ms}. NAS is becoming increasingly successful mainly because it liberates human designing labors by automatically identifying high-performance neural architectures. Co-exploration of neural architecture and hardware design~\cite{jiang2019accuracy, jiang2020standing, jiang2020hardware} push this concept further by incorporating hardware design specifications into NAS search spaces, so as to offer neural architecture-hardware design pairs that are accurate, efficient, and robust against hardware uncertainties.

Formally speaking, NAS deals with a problem that, given a perceptual task $T$, a human-defined design space $\mathbf{S}$, and a set of figures of merit (FOM) $\mathbf{P}$, what is the best neural architecture in $\mathbf{S}$ that can offer optimal performance (in terms of FOM in $\mathbf{P}$) on task $T$.

A typical reinforcement learning (RL)-based NAS that solves this issue, such as the framework proposed in~\cite{zoph2016neural}, is composed of three key components, a \emph{controller}, a \emph{trainer} and an \emph{evaluator}. In one iteration (named \emph{episode}) of RL-based NAS, (1) the \emph{controller} generates a neural architecture from the design space; (2) the \emph{trainer} builds the generated neural architecture into a DNN model, named child network, and trains the child network on a held-out training dataset; (3) the \emph{evaluator} collects the figures of merit (FOM), \emph{e.g.}, test accuracy of the trained child network on test dataset, its latency and/or energy consumption; and (4) the \emph{controller} use a user-defined reward function to calculate a $reward$ data from FOM collected by the \emph{evaluator} and use the $reward$ to update itself so that it can predict neural architectures with higher FOM.

This iterative method terminates under two circumstances: (1) the \emph{controller} repeatedly predicts the same child network; and (2) the number of predicted architectures exceeds a predefined threshold (episode limit). The child network that offers the highest \emph{reward} among all the generated neural architectures is presented as the search result. The chosen neural architecture is then re-trained on the training dataset for a longer training time to offer optimal performance.

More recently, differentiable NAS~\cite{liu2018darts, wu2019fbnet, li2020edd} has achieved state-of-the-art performance with a much-reduced search time by transforming the search process into training an over-parameterized neural network. However, those approaches suffer from flexibility. More specifically, in the field of research considered in this paper, differentiable NAS struggles in handling large search spaces with multiple different hardware design parameters and complex designs where the number of channels varies for each layer. Thus, in this work, we adopt RL based NAS as our search algorithm.

\section{\todo{Uncertainty Modeling}}\label{Sect:Analysis}
\subsection{Uncertainty Model}
In this work, we model device uncertainties as a whole and use a Gaussian distribution to represent them~\cite{jiang2020device, feinberg2018making}. We set the mean of the uncertainty distribution to be zero, its variation to be 0.04, and for each device, its uncertainty is independently distributed. which is referred from~\cite{zhao2017investigation}, where the uncertainties are measured from actual physical devices. For an easier representation of the latter part of this paper, the uncertainty model is depicted as:

\begin{equation}\label{eq:noise}
    W_{Dep} = W_{Exp} + \mathbf{N}(\mu, \sigma)
\end{equation}
where N is a Gaussian variable which, on each individual element of the weights, is independent and identically distributed. $W_{Exp}$ and $W_{Dep}$ are the expected weights trained in the data center and the actual weights deployed on the accelerators, respectively.

\subsection{Effects on DNN Outputs}\label{sec:changeMethod}

In this work, we focus on the impact of device uncertainty on classification tasks, and the reasons go as follows: (1) most emerging device-based DNN accelerators target classification tasks, analyzing the effect of device uncertainties on these tasks helps the majority of the researchers to improve their work; (2) DNNs for classification tasks are typically composed of convolution layers and fully connected layers, which are also the basic components of DNNs targeting other application. The effect of device uncertainties on these two components is essential for all types of DNNs.

We start by understanding the effect of device uncertainties on the output of a DNN model. Formally speaking, a DNN model $M$ can be defined as a combination of its neural architecture and its trained weights. Thus, the inference process of a DNN model with input $\mathbf{I}$ that generate an output $\mathbf{O}$ can be defined as:

\begin{equation}\label{eq:NN}
    \mathbf{O} = F\ (W, \mathbf{I})
\end{equation}
where $F$ is the neural architecture of $M$, $W$ is its weights, $\mathbf{I}$ is this input vector and $\mathbf{O}$ is the output vector. 

During training, $\mathbf{O}$ is then passed through a loss function, where a version of $\mathbf{O}$ after SoftMax is compared with the ground truth classification label $GT$ to generate a \emph{loss} for the backpropagation process. During inference, the final predicted class of $\mathbf{I}$ can be calculated by $argmax(\mathbf{O})$, which is the index of the item in $\mathbf{O}$ that has the maximum value.

Although in inference, classification result is the final outcome of a DNN model, the output vector $\mathbf{O}$ serves as a better representative of the behavior of this model. The classification result is only an index of the maximum value of $\mathbf{O}$ and is thus only a simplified discrete proxy of $\mathbf{O}$. The continuous, multi-dimension vector $\mathbf{O}$ contains more information than the classification result. In order to understand how uncertainties in weights may affect the network, it is of crucial importance to understand how it affects $\mathbf{O}$.

As defined in~\ref{eq:noise} and ~\ref{eq:NN}, a deployed neural network under the effect of device uncertainties can be depicted as:
\begin{equation}\label{eq:deploy}
    \mathbf{O}_{Dep} = F\ (W_{Dep},\ I) = F\ (W_{Exp} + N_j,\ I) 
\end{equation}
where $W_{Exp}$ is the trained value of the neural network to be deployed, $N_j$ is one sample from the noise distribution, and $\mathbf{O}_{Dep}$ is the affected output.

We analyze the distributional behavior of the effect of device uncertainties on the output vector of a DNN model. To conduct this analysis, we first (1) train a DNN model $F$ to converge and collects its trained weight $W_{Exp}$. We then (2) fix one input image $I$ and collect its output on the trained weight $W_{Exp}$. We denote this output as the original output $\mathbf{O}_{Ori}$. After that, we (3) sample $K$ different instances of noises $N_1, N_2, ... N_K$ and then feed them to Eq.~\ref{eq:deploy}, collecting $K$ different output vectors. Finally, we calculate \emph{output change} using Eq.~\ref{eq:change}

\begin{equation}\label{eq:change}
    \mathbf{O}_{CG} = \mathbf{O}_{Dep} - \mathbf{O}_{Ori}
\end{equation}
where the output change is the element-wise subtraction of the perturbed and original output.

\subsection{Experimental Results}

In order to get a glance at the statistical behavior of \emph{output change}, according to the workflow introduced in Sect.~\ref{sec:changeMethod}, we train a LeNet model for the MNIST dataset~\cite{lecun1998gradient} to state-of-the-art accuracy. We then randomly choose one input image in the test dataset and sampled 10k different instances of noise. with this setup, we gathered 10k different \emph{output change} vectors.

The \emph{output change} is a vector of 10, with each element representing the confidence of classifying the input image into one certain category. Because a high-dimensional vector is not a good choice for analytical study and visualization, we analyze each element of these vectors. We analyze the statistical behavior of each element across different vectors and gathered 10 instances of distribution data.

\begin{figure}[h]
\begin{center}
\centerline{\includegraphics[trim=0 390 0 0, clip, width=0.6\linewidth] {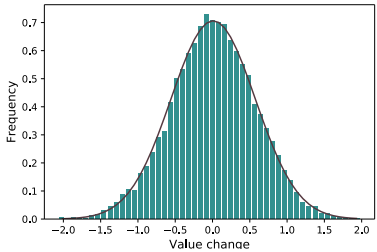}}
\caption{\emph{Output change} distribution of LeNet for MNIST. 10k \emph{output change} vectors are gathered from one trained LeNet model affected by 10k different instances of noise sampled from $\mathcal{N}(0, 0.04)$. This figure shows the distribution of the fist item of the gathered \emph{output change} vectors.}
\label{fig:MNIST_Output}
\vspace{-0.5cm}
\end{center}
\end{figure}

Surprisingly, each element of the \emph{output change} follows Gaussian distribution. To visualize this finding, we plot the histogram of the distribution of each element of \emph{output change} vector and the corresponding Gaussian distribution that fits it. The visualization result for the first element of \emph{output change} is shown in Fig.~\ref{fig:MNIST_Output} that the first element of \emph{output change} vector is a Gaussian variable.

To verify this observation, We tested various networks in various datasets. With the MNIST dataset, we analyze both LeNet and multilayer perceptrons (MLP) using ReLU and Sigmoid activation with 2 layers. With the CIFAR-10 dataset~\cite{krizhevsky2009learning}, we test a conventional floating-point CNN, a quantized CNN, and two ResNets, ResNet-56 and ResNet-110. We also train these models with 3 different initializations to get different trained weights. 

We evaluate how these \emph{output change} vectors fit into Gaussian variables by two widely used standards: mean square error (MSE) and Chi-square ($\chi^2$) test. MSE can be described as:

\begin{equation}\label{eq:chi2}
    MSE = \frac{1}{N} \sum^{N}_{i=1}(O_i - E_i)^2
\end{equation}
and $\chi^2$ test can be depicted as:
\begin{equation}\label{eq:chi2}
    \chi^2 = \sum^{N}_{i=1}\frac{(O_i - E_i)^2}{E_i}
\end{equation}
where $O_i$ and $E_i$ are the observed (\emph{output change}) and estimated (Gaussian) value of probability and $N$ is a user-defined granularity. We define $N = 100$ because it is precise enough when we have a total of 10k instances of data.

\begin{table}[h]
    \centering
    \begin{tabular}{lccc}
        \toprule
        Model & Dataset & $\chi^2$ ($10^{-2}$)& MSE ($10^{-4}$)\\
        \midrule
        MLP-ReLU    & MNIST     & 5.22 & 3.20\\
        MLP-Sigmoid & MNIST     & 5.81 & 2.20\\
        LeNet       & MNIST     & 4.59 & 2.67\\
        Float-Conv  & CIFAR-10  & 7.01 & 3.03\\
        Fixed-Conv  & CIFAR-10  & 6.79 & 2.74\\
        ResNet-56   & CIFAR-10  & 4.56 & 1.79\\
        ResNet-110  & CIFAR-10  & 4.81 & 2.01\\
        \bottomrule
    \end{tabular}
    \caption{Gaussian fit for different models. The $\chi^2$ test result and MSE between the \emph{output change} and Gaussian distribution is presented. Both tests show that the \emph{output change} follows Gaussian distribution w.r.t different instances of noise.}
    \label{tab:fit}
    \vspace{-0.5cm}
\end{table}

The evaluation result is shown in Table~\ref{tab:fit}. Note that we have test 3 different initializations for each model and for each model, $output change$ is a vector of 10. The result shown in Table~\ref{tab:fit} is an average of them.
For each model tested, $\chi^2$ test results are all below 0.1 and MSE are all below $1\times 10^{-3}$, which indicates that they are well fit into Gaussian distributions. Moreover, both errors do not increase when the model is extremely shallow (\emph{e.g.} 2-layer MLP) and very deep (RestNet-110), so this observation generalizes across different DNN models.

The study on each of the models supports the previous observation that their output vectors values follow Gaussian distribution. Based on these studies, we can claim that,

\noindent\textbf{with any independent and identically distributed Gaussian noise on weight, the output vector of the same input image follows a multi-dimensional Gaussian distribution\footnote{Note that each element of the output are deeply co-related, not independent} over different samples of noise.}

This is a very strong claim but is not counter-intuitive. The output of the first convolution layer is the summation of the multiplication result of deterministic inputs and Gaussianly distributed weights and is thus a summation of Gaussian distributions. The summation of Gaussian variables is also a Gaussian variable, so the output of the first layer is a Gaussian variable. After activation, the input of the second layer is a transformed Gaussian variable and after propagating through this layer, with enough number of operands, the accumulated variable can be approximated by Gaussian variables. Thus, although the final output may not strictly be a Gaussian variable, a Gaussian approximation can be observed.




\section{Uncertainty Aware Search}

\subsection{Methodology}
In addition to understanding the effect of device uncertainties, we propose a remedy method to reduce the effect of this issue by adopting NAS. 

In this work, we propose \underline{U}ncertainty \underline{A}ware s\underline{E}arch (UAE), a more comprehensive uncertainty aware NAS for better exploration of neural architectures in non-ideal emerging devices based CiM NN accelerators. 

Similar to the state-of-the-art Reinforcement Learning based NAS framework NACIM~\cite{jiang2020device}, UAE works iteratively and in each iteration: (1) an LSTM-based \emph{controller} is used to learn and generate neural architectures; (2) an uncertainty aware \emph{trainer} is used to train each generated neural architecture to get a model to deploy; (3) an uncertainty aware \emph{evaluator} is adopted to evaluate the actual performance of the deployed model; (4) the evaluated performance is used as a reward to update the \emph{controller} so that it can generate neural architectures with higher rewards.

The detailed implementation of the uncertainty aware \emph{trainer} and \emph{evaluator} are described below.

\subsection{Uncertainty-Aware Training \& Evaluation}\label{sec:train}

We adopt an uncertainty-aware training scheme similar to the one used in NACIM~\cite{jiang2020device}. The training process is organized the same as traditional DNN training that, in each iteration, a subset (batch) of the training data is used to train the model, and after the whole training dataset has been used to train the model, and an \emph{epoch} of training is finished and another \emph{epoch} is started. The \emph{trainer} trains the model for multiple \emph{epochs} to get a trained model.

The uncertainty-aware training augments the training process for each batch to learn a DNN model that is more robust against device variations. In each training batch, before feeding the input into the model, the \emph{trainer} (1) save the original weight $W_{Ori}$ of the model; (2) sample a noise from the uncertainty distribution and add the noise to the weight of the model to form a $W_{Dep}$; (3) perform forward inference and back propagation in the perturbed model and collect gradient data for each weight; (4) load the saved $W_{Ori}$ back to the model and update $W_{Ori}$ with the collected gradient data via stochastic gradient descent.

Uncertainty-aware training simulates the process of training DNNs directly on CiM-based accelerators. Experiments in NACIM show that uncertainty-aware training learns DNN models that are robust against device uncertainties.

The uncertainty-aware evaluation is performed similarly to the training process. Before evaluation, the \emph{evaluator} samples an instance of noise from the uncertainty distribution and add the noise to the trained weight $W_{Ori}$ to get a $W_{Dep}$. The \emph{evaluator} then evaluates the classification accuracy of the perturbed model on a test dataset. This process is performed for $K$ times and $K$ different accuracy data are gathered. \todo{The \emph{evaluator} then report one distributional property (\emph{e.g.}, mean, maximum value, 95\% minimum value) of the $K$ accuracy data to form a \emph{reward}}. The distributional property to be used is specified by the user. 

\subsection{Experimental Results}

We demonstrate the effectiveness of UAE by searching for an optimal quantized CNN for CIFAR-10. The fixed design parameters and hyper-parameters included in the search space are shown in Table~\ref{tab:space}. For device uncertainty specifications, we assume an ISAAC-like~\cite{shafiee2016isaac} neural accelerator architecture and a four-bit RRAM device, whose behavioral model is extracted from~\cite{zhao2017investigation}. The search process is conducted in a GPU server machine with an Nvidia GTX 1080ti accelerator.

\begin{table}[h]
    \centering
    \begin{tabular}{lcl}
        \toprule
        Hyper-Parameters    && Value choices \\
        \midrule
        Dataset             && CIFAR-10 \\
        Type                && Quantized CNN \\
        \# of Conv Layers   && 6 \\
        \# of FC Layers     && 2 \\
        FC Hidden size      && 1024\\
        \midrule
        \# channels         && (24, 36, 48, 64) \\
        Filter Height/Width && (1, 3, 5, 7) \\
        \# of integer bits  && (0, 1, 2, 3) \\
        \# of fraction bits && (0, 1, 2, 3, 4, 5, 6) \\
        \bottomrule
    \end{tabular}
    \caption{Quantized CNN for CIFAR-10 search setups. Upper-half: configurations  fixed to be the same among all searched architectures; lower half: hyper-parameters to be searched.}
    \vspace{-0.5cm}
    \label{tab:space}
\end{table}

As described in Sect.~\ref{sec:train}, there are two major search parameters: the instances ($K$) of noise sampled for each architecture and the distributional properties used to form the accuracy data collected by the \emph{evaluator} into a \emph{reward}. We test two different values of $K$, 5, and 100 samples, for the reason that will be explained afterward. We also use two different distributional properties, one is the mean value of all accuracy data (mean) and the other is 95\% minimum of the accuracy data (95). The mean value indicates how a model performs under the effect of device uncertainty in average circumstances and the 95\% minimum shows the models' behavior in worst-case scenarios.


We offer a comparison for different specifications of UAE and two baseline methods, quantNAS~\cite{Lu2019Neural}, a state-of-the-art NAS framework to search for the optimal quantized CNN and NACIM~\cite{jiang2020device}, another uncertainty aware searching framework for CiM-based accelerators. In each experiment, the NAS \emph{controller} searches for 2000 different architectures (episodes) and the \emph{trainer} trains each generated architecture for 15 epochs.

The DNN models finally presented by each search framework are also evaluated by mean and 95\% minimum value with their accuracy data collected by 10k Monte-Carol simulation. The experimental result is shown in Table.~\ref{tab:result}.\footnote{Because the data for quantNAS and NACIM are collected from published work, we do not have the 95\% minimum accuracy result for them.}

\begin{table}[h]
    \centering
    \begin{tabular}{cccccc}
        \toprule
        Method & $K$ & w/o noise & mean & 95 & Time (h)\\
        \midrule
        QuantNAS~~\cite{Lu2019Neural} & 0  & 84.92\% & 08.48\% & N/A & 53\\
        NACIM~\cite{jiang2020device}   & 1   & 73.88\% & 73.45\% & N/A & 98\\
        UAE-M   & 5   & 77.48\% & 75.94\% & 75.55\% & 118\\
        UAE-M   & 100 & 82.99\% & 79.84\% & 77.82\% & 255\\
        UAE-95  & 100 & 80.64\% & 78.39\% & 77.98\% & 255\\
        \bottomrule
    \end{tabular}
    \caption{Comparison for different specifications of UAE and two other baselines. Different methods sample different instances of noise ($K$) in uncertainty-aware evaluation. UAE-M uses mean value of the accuracy data collected by the \emph{evaluator} to form a reward and UAE-95 uses the 95\% minimum data. Accuracy without noise shows the model accuracy with an ideal device and the mean and 95\% minimum (95) value shows the behavior of the searched DNN models under the effect of device uncertainty, evaluated by Monte-Carol simulation.}
    \label{tab:result}
    \vspace{-0.5cm}
\end{table}

Experimental results show that, without uncertainty-aware training, QuantNAS can identify an optimal DNN model that can offer close to 85\% of test accuracy, but struggles in finding proper neural architectures that are robust to device uncertainties, as the test accuracy of the DNN model identified by quantNAS is down to 8.5\%, even worse than random guessing (10\%). With the help of uncertainty-aware training, NACIM can identify DNN models that are robust against the impact of device uncertainties. However, as NACIM only evaluates the performance of the searched architecture once, the randomness of the device uncertainty hinders the performance of NACIM, resulting in a model with only 73.45\% of average accuracy. UAE, on the contrary, is able to find DNN models that are both accurate and robust against the impact of device uncertainties. When collecting only 5 samples in uncertainty-aware evaluation, UAE achieves 2.49\% higher accuracy than NACIM with a time overhead of only 20\%. When collecting 100 samples in uncertainty-aware evaluation, UAE achieves 6.39\% higher accuracy than NACIM a search time overhead of 2.5x. Though further increasing the number of samples is possible, the search time overhead will be too large to endure. The adoption of a 95\% minimum value standard is also effective. UAE-95 offers 0.16\% higher worst-case accuracy than its UAE-M counterpart with a 1.45\% lower average accuracy.

\section{Conclusions and Future Works}

In this work, we propose a Monte-Carlo simulation-based experimental flow to measure the device uncertainty-induced perturbations to DNN models. Our thorough investigation of the behaviors of different DNN models under such perturbations and shows that the value changes of their output vectors follow Gaussian distribution. We also propose UAE, a device uncertainty-aware NAS framework that identifies DNN models that are both accurate and robust against device uncertainty-induced perturbations.
Based on the observations made on the impact of device uncertainties on the DNN models, the possible future directions include the formal mathematical proof of the analyzed statistical behaviors and a time-efficient estimation method for the impact of device uncertainties.

\section*{Acknowledgment}

This work is supported in part by National Science Foundation under grant CNS-1919167.

\bibliographystyle{ACM-Reference-Format}
\bibliography{M7_References.bib}

\end{document}